\documentclass[10pt,a4paper]{article}
\usepackage{amsmath}
\usepackage{latexsym}
\usepackage[pdftex]{graphicx}
\usepackage[utf8]{inputenc}
\usepackage{authblk}

\title{Can we observe the gravitational quantum states of Positronium?}
\author[1]{P. Crivelli\thanks{crivelli@phys.ethz.ch}}
\author[2]{V. V. Nesvizhevsky}
\author[3]{A. Yu. Voronin}

\affil[1]{\small{ETH Zurich, Institute for Particle Physics, Otto-Stern-Weg 5, 8093 Zurich, Switzerland.}}
\affil[2]{\small{Institut Laue-Langevin, 6 rue Jules Horowitz, Grenoble, F-38046, France.}}
\affil[3]{\small{Lebedev Physical Institute, 53 Leninsky prospect, Moscow, 119991, Russia.}}
\date{}

\begin{document}

\maketitle
\begin{abstract}
In this paper we consider the feasibility of observing the gravitational quantum states of positronium. The proposed scheme employs the flow-throw technique used for the first observation of this effect with neutrons. Collimation and Stark deceleration of Rydberg positronium atoms allow to select the required velocity class. If this experiment could be realized with positronium it would lead to a determination of $g$ for this matter-antimatter system at the few \% level. As discussed in this contribution, most of the required techniques are currently available but important milestones have to be demonstrated experimentally before such an experiment could become reality. Those are: the efficient focusing of a bunched positron beam, Stark deceleration of Rydberg positronium and its subsequent excitation into states with large angular momentum. We provide an estimate of the efficiencies we expect for these steps and assuming those could be confirmed we calculate the signal rate. 

\end{abstract}

\section{Introduction}
Quantum gravitational states were observed for the first time with neutrons by measuring their transmission through a slit made of a mirror and an absorber \cite{valery2002}. If the distance between the mirror and the absorber (which is a rough surface used as a scatterer to mix the velocity components) is much higher than the turning point for the corresponding gravitational quantum state, the neutrons pass through the slit without significant losses. As the slit size decreases the absorber starts approaching the size of the neutron wave function and the probability of neutron loss increases. If the slit size is smaller than the characteristic size of the neutron wave function in the lowest quantum state, the slit is not transparent for neutrons as this was demonstrated experimentally.
The height and energies of the gravitational quantum states can be determined analytically and the solution of the Schroedinger equation contains Airy functions. A more transparent and simple equation can be derived using a semi-classical approach \cite{QMbooks}. This solution reproduces the energy of the gravitational states within 1\% and is given by: 
\begin{equation}\label{eq:en}
E_n\simeq \sqrt[3]{\frac{9 m}{8}\cdot(\pi \hbar g (n-\frac{1}{4}))},
\end{equation}
where $m$ is the particle mass, $g$ the gravitational acceleration, $n$ the principal quantum number and $\hbar$ the reduced Planck constant.
The characteristic scale for the gravitational quantum states is equal to:
\begin{equation}\label{eq:z0}
z_0 = \sqrt[3]{\frac{\hbar^2}{2 m^2 g}}.
\end{equation}

The corresponding classically allowed heights are given by:
\begin{equation}\label{eq:z1}
z_n = E_n/mg = \lambda_n z_0. 
\end{equation}
where $\lambda_n=\{2.34,4.09, 5.52, 6.79, 7.94, 9.02, 10.04...\}$ are the zeros of the Airy-function.
For neutrons the height of the lowest gravitational level is 13.7 $\mu$m. For positronium, the electron-positron bound state, that is 1000 times lighter than a neutron one gets a 100 larger size corresponding to $z_1 =1.3$ mm while the energy is 10 times smaller, $E_1=0.13$ peV. The observation time to resolve a quantum gravitational state can be estimated using the Heisenberg uncertainty principle to be of the order of $\hbar/E_1 \simeq 4.5$ ms.
This value is much larger than the long lived triplet positronium lifetime in the ground state which is 142 ns (the Ps singlet state lives only 125 ps and thus in the following we will only consider the triplet state and refer to it as Ps). Luckily, the Ps lifetime can be increased by excitating to a higher level. In a Rydberg state the Ps lifetime against annihilation is increased by a factor of $n^3$, where $n$ is the principal quantum number, because of the decrease of the overlap of the positron and the electron wave functions.
As for the case of a measurement of the gravitational free fall of Ps proposed by Mills and Leventhal \cite{mills2002}, the usable lifetime to observe a quantum gravitation state of a Rydberg Ps atom (hereafter Ps$^*$), is the one before it emits the first photon. In fact, after that the recoil will modify its trajectory and vertical energy thus the Ps atom will be lost inside the slit. 
For this reason, the excited Ps has to be spun up to high $l$ values with circularly polarized microwave radiation.

\section{Experimental technique}
\begin{figure}
\begin{center}
\includegraphics[width=1.\textwidth]{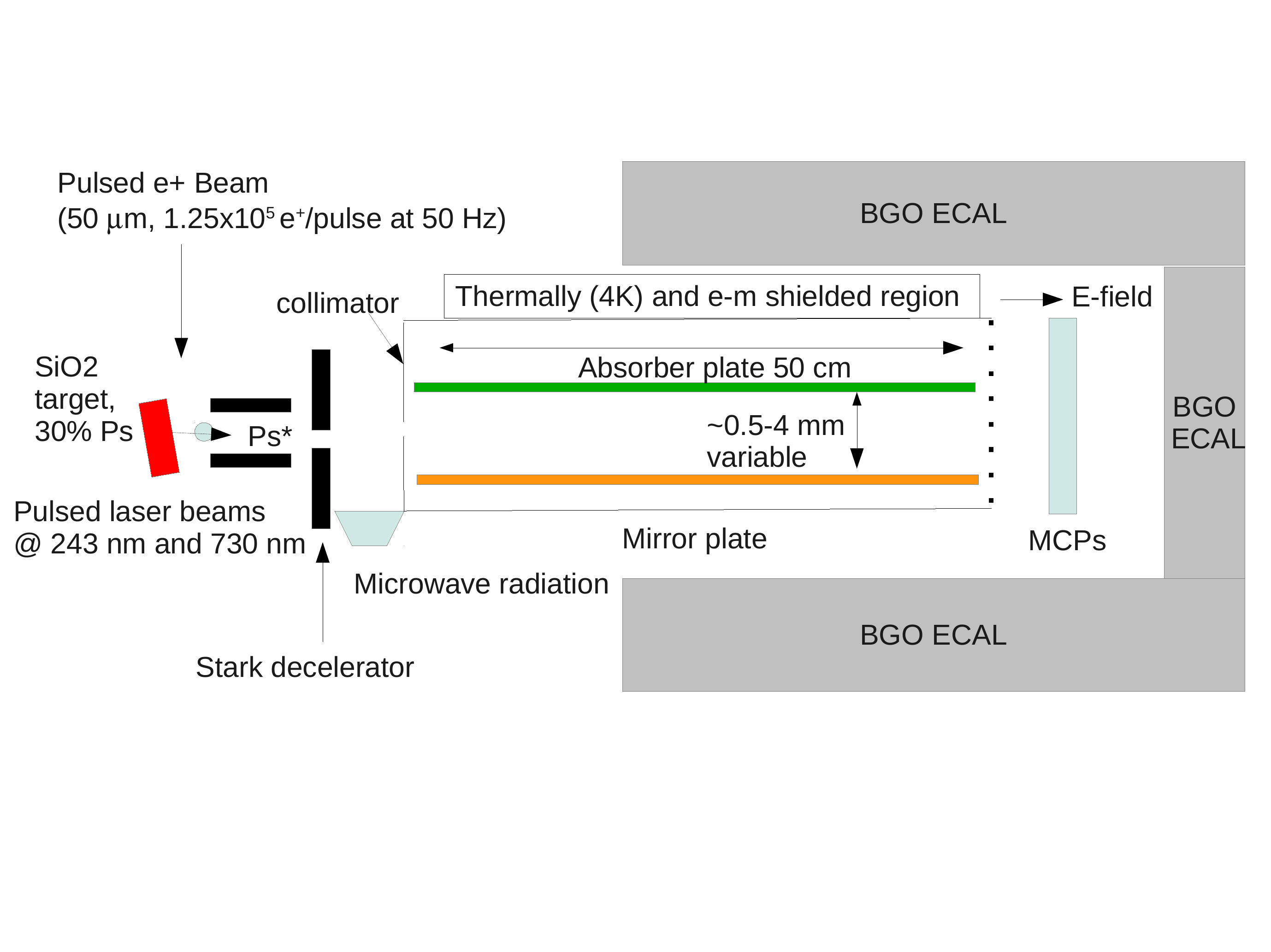}
\caption{\em Scheme of the proposed experimental setup.}
\label{fig:scheme}
\end{center}
\end{figure}

A scheme of the proposed experimental set-up is shown in Fig. \ref{fig:scheme}. Positronium is formed by implanting keV positrons from a re-moderated pulsed slow positron beam in a positron-positronium converter (see Section \ref{target}).
To observe the quantum mechanical behavior of Ps in the gravitational field its vertical velocity should be of the same order as the gravitational energy levels and thus $v_y<0.15$ m/s. Furthermore to resolve the quantum state the Ps atom has to interact long enough with the slit and therefore it has to be laser excited to a Rydberg state with $n>30$ and maximum $l$ quantum number. To keep a reasonable size of the experimental setup (i.e a slit size of the order of 0.5 m) and minimize the number of detectors the velocities in the horizontal plane should be smaller than $v_{x,z}<100$ m/s. Similar to neutrons a collimator will be used to select the velocity components $v_x,v_y$ of the positronium distribution. However since no reliable thermal cold source of positronium exists the velocity component perpendicular to the surface $v_z$ has to be lowered by some other means. Relying to the fact that atoms in Rydberg states have a large dipole moment Stark deceleration can be used for this purpose.  
After deceleration the Ps$^*$ are driven by circularly polarized microwaves to a state with high $l$. If the slit width is smaller than the first expected gravitational state (i.e $<1$ mm) this will not be transparent and therefore no signal will be detected above the expected background in the detectors. If the width is increased to a value lying between the first and the second gravitational state (i.e. $<2$ mm) the Ps wavefunction can propagate and a signal is expected to be detected. This quantum jump provides the unambiguous indication of the observation of a quantum gravitational state of positronium.

\section{Experimental setup}

\subsection{Positron beam}

Few facilities around the world are nowadays capable to produce fluxes of positrons exceeding $1\times10^8$ e$^+$/s \cite{SlowPosBeams}. The highest intensity reported so far has been reached at the FMR II NEMOPUC source in Munich where a mono-energetic slow positron beam of $9\times 10^8$ e$^+$/s has been achieved \cite{NEMOPUC}.
We will base our estimation of the signal rate assuming that such a beam could be used. In principle, even higher fluxes are conceivable (see e.g. \cite{APosS,JLab}) and hopefully those will be available in the near future.  

The beam will be coupled to a buffer gas trap. This is necessary in order to produce positron pulses at a rate of tens of Hz to synchronize positronium formation with pulsed lasers in order to drive the transition of Ps in a Rydberg state. The typical trapping efficiency is 20\% and the positron lifetime in the trap is 2 s. Therefore above a few Hz repetition rate the losses are minimized. The positrons are dumped in bunches of $20$ ns using the standard electrodes of the trap and by adding a buncher one can achieve $<$1 ns pulses \cite{cassidy2006,beamNIMA}. 
The typical positron beam size of such a system is 1 mm (FWHM). This has to be reduced via positron re-moderation in order to achieve a good geometrical overlap of the Ps source with the opening of the slit (mirror-absorber). 

To note is that a new concept for a positron buffer gas trap with the prospect of obtaining up to 92\% trapping efficiencies has been recently proposed \cite{dan}.

\subsection{Re-moderation}\label{mod}
Re-moderation of positron beam allows brightness enhancement. Micro-beams of 1-2 $\mu m$ were developed by the use of two successive stages of focusing, acceleration and re-moderation \cite{brandes}. The loss of beam intensity is about a factor 10 per stage in reflection geometry. To achieve a positron beam of 50 $\mu m$ only one moderator stage will be required. One can use a Ni(100) foil in transmission geometry which even being a factor 2 less efficient allows the beam to be extracted from the magnetic field region with a transmission of 71\% \cite{Ni100}. Such a re-moderator leads to about 5\% of the incoming positrons in a beam of 50 $\mu m$ \cite{Ni100}. This was achieved using a continuous positron beam but it has not been yet demonstrated for a bunched beam. Such a scheme is currently under investigation at the University College of London (UCL) for their ongoing free fall measurement of Ps. Their preliminary simulations show that one should be able to obtain 1 ns positron bunches of about 25 $\mu m$ spot size \cite{Psfreefall}. Assuming that a remoderation efficiency of 5\% could be attained, the positron flux at the target will be of the order of $6.3\times10^6$ e$^+$/s ($1.25\times10^5$  e$^+$ per bunch at a repetition rate of 50 Hz).  
 
\subsection{Positronium source}\label{target}
Positrons implanted in silica thin films with 3 keV will form Ps emitted into vacuum with an efficiency of about 30\% and an energy of 75 meV \cite{crivelliPRA,cassidyPRA}.  The vacuum yield is basically constant in the temperature range from 50 K to 400 K \cite{crivelliPRA}. The emission energy of Ps is determined by the diameter of the pores (3 nm). In fact for Ps energies of a few hundred meV the Ps de Broglie wavelength is comparable with the energy levels of positronium confined inside a pore of this size and therefore it has to be treated quantum mechanically. A simple particle in a box model is a good approximation to describe Ps in such porous materials \cite{gidley,crivelliPRA,crivelliPRB}. Ps formed inside the film with an initial energy of 1 eV (corresponding to the negative workfunction of Ps in silica) start diffusing through the interconnected pore network tunneling from one pore to the other and looses energy via collision until the ground state energy is attained. A fraction of those atoms (30\% of the incoming positrons) will make it to the surface before decaying and are emitted into vacuum. At temperatures below 100 K in films of 3-4 nm diameter only the ground state of Ps in the pore is populated. Therefore, Ps is emitted into vacuum is basically mono-energetic. Since the pore orientation at the surface of these materials is random the emission distribution follows a $cos\theta$ distribution \cite{crivelliPRA,cassidyPRA}.

 At 3 keV the positrons have a mean implantation depth of 150 nm. The time required for the Ps atoms to exit the film and be emitted into vacuum has been measured to be about 1 ns \cite{cassidyDelayed}. Since the areal density per pulse is of the order of $10^{9}$ Ps/cm$^{2}$ the spin exchange quenching and formation of molecular positronium inside the films should be considered. The total cross section in those targets for these processes has been determined to be $3.4 \times 10^{-14}$ cm$^{-2}$ \cite{CassidyPsPs}. In those measurements a source based positron beam was used and at areal densities of $4\times 10^{10}$ Ps/cm$^{2}$ a saturation of the quenching process was observed. A fraction of 0.3 of the initial positronium population survives since the positrons from a radioactive source are spin polarized (due to parity violation). For positrons produced in a reactor this is not the case and therefore the losses will be higher. However, using the measured cross sections and taking into account that once emitted into vacuum after 1 ns the Ps cloud expands and therefore the density decreases rapidly we estimated that this effect can be neglected.

 To reduce the perpendicular velocity $v_z\simeq10^5$ m/s we plan to use Stark deceleration. This method has been applied to different atomic species (including hydrogen) \cite{merkt2008} and molecules \cite{merkt2009}. Atoms in Rydberg states have large dipole moments thus electric field gradients can be used to manipulate them. The acceleration/deceleration $a$ imparted to the Rydberg atoms is given by:  
\begin{equation}
a=76 \nabla F \frac{1}{m} n k,
\end{equation}
where $\nabla F$ is the gradient of the electric field in Vcm$^{-2}$, $m$ the mass of the decelerated particles in atomic units, $n$ and $k$ the Stark state quantum numbers. H atoms in $n=25$ and an initial velocity of 700 m/s can be brought at rest in 3 mm  \cite{merkt2008}.
For Ps being 1000 times lighter decelerations exceeding $10^9$ m/s could be realized. Since one is interested only in decelerating the distribution that is almost perpendicular to the surface of the Ps target, one can expect an efficiency close to 100 \%. This is confirmed by preliminary simulations \cite{seiler}. 
The collimator will be placed after the deceleration stage and the microwave region where circularly polarized radiation will spun up the Ps to the maximum $l$ so that kicks to the momentum imparted to the atoms in the vertical direction during these process will be accounted for. 

 The fraction of atoms with $v_y<0.15, v_x,v_z<100$ m/s is estimated using a Monte Carlo simulation taking into account the angular of the Ps emitted into vacuum from porous silica to be of the order of $2 \times 10^{-9}$.

Other promising candidates for Ps formation that should be considered for this experiment are aluminum oxidized nano-channels of 5-8 nm \cite{brusaPRL}, Si and Ge surfaces \cite{CassidyPRB2012} and hydrophobic silica aerogels \cite{Consolati}.

To produce a source of cold Ps another option would be to use the tails of the distribution of Ps thermally desorbed and emitted into vacuum. For Al(111) kept at 600 K the conversion efficiency e$^+$-Ps was measured to be as high as 30\%  \cite{millsAl111}. However, the surface quality is very important, and degradation is observed in a short time scale especially when a laser is positioned near the surface \cite{Fee:1993zz,Ps2}. Therefore, until stable desorption at room temperature (or lower) could be demonstrated this does not seem to be an attractive path.

\subsection{Laser for Rydberg excitation} 
Positronium excitation in Rydberg states was first demonstrated by Ziock et al \cite{ziock}. Recently, an excitation of Rydberg Ps with a two step process 1S$\to$2P and 2P$\to$25 with an efficiency of 25\% was reported \cite{cassidyRydberg,jones}. Two broad band (150 GHz) Dye lasers were used in order to maximize the overlap with the large Doppler profile of the Ps atoms. This led to an occupation of the Stark levels that could not be resolved. 
In the experiment proposed here one is interested to excite only a small fraction of the atoms with the correct velocity characteristic. The lasers axis will be set parallel to the $v_x$ component. For $v_x=100$ m/s the Doppler profile will be of the order of 1 GHz and thus narrow dye lasers at 243 nm and 730 nm can be used. One has to consider that in both cases the photon absorbed by the atom will result in a recoil and therefore the laser frequency has to be tuned in order to compensate for this effect. This can be optimized by detecting the Rydberg atoms via field ionization after the collimator slit maximizing their flux.  
In such a scheme the excitation probability will be at least as large as 25\% (as for the case of broadband excitation) with the advantage of populating only the desired Stark state ($n=33$ and $k=19$). As a base for our estimation of the signal rate we take conservatively this value.
An alternative, might be to use Doppler free two photon excitation from the ground state directly to $n=33$. This could lead to higher efficiencies exceeding 50\% \cite{Psfreefall,WAG2013} but it has not been yet realized for Ps. 
After deceleration, the Ps$^*$ will be driven to its highest circular state $l=32$ using polarized microwave radiation.  

\subsection{Mirror and absorber} 
As a mirror for Ps we propose to exploit a gradient of magnetic field created using wires arranged parallel to each other with a constant current to create a uniform gradient of the magnetic field. Only the Ps triplet atoms with $m=0$ have a non-zero net magnetic moment. For the $m=\pm1$ the electron and the positron magnetic moments cancel and therefore those are insensitive to the magnetic field. Therefore, only one third of the initial population will be reflected. 
To equate the $E_y = 0.1$ peV a field of a few mG at the wire surface will be sufficient.

Because of the large spacial size of gravitational quantum states and the very large characteristic length of the mirror needed to form the gravitational states that is  much larger than a characteristic inter-wire distance,  we expect that the very weak magnetic gradient will not perturb the gravitational states.
The strict theoretical analysis of this clearly mathematically defined problem is ongoing.

It is important to note that a well defined $(n,l)$ state should be used in order to avoid that the spread in the magnetic moment would wash out the transmission versus the slit height dependence.

A matter mirror should also be considered. Due to the large spacial size of the gravitational quantum state,  the surface potential is expected to be very sharp and therefore result in efficient quantum reflection \cite{dufour}.
In both cases (magnetic or material mirror) we expect to have
effectively (quasi-classically) only a few collisions with surface. Nevertheless, the transitions rates due to quenching and ionization caused by the electric or magnetic fields have to be calculated.

The absorber as for the neutrons is a rough surface on which the impinging Ps will be mix their velocity components be lost.

\subsection{Detectors} 
The detectors is made of an array of ten standard rectangular MCPs (e.g. Hamamatsu F4772-01, 55x8 mm$^2$) placed 50 mm from the end of the mirror-absorber region. A uniform electric field will ionize the Ps$^*$ and guide the positrons to the MCPs where they can be detected with high efficiency (80\%). In order to reduce accidental background, the MCPs will be surrounded by an almost 4$\pi$ calorimeter made of 100 BGOs (hexagonal shape of 55 mm outer diameter and 200 mm length) placed outside of the vacuum chamber to detect the back to back 511 keV annihilation gamma rays from the e$^+$ annihilation. This detector was used in the search for Ps$\to$invisible decays were the efficiency for detection of two 511 keV photons was larger than 99.99999$\%$ \cite{oPsInv}. 
 An energy cut of 400 keV $<E<$ 600 keV will be applied and the coincidence between the two back to back crystals within 10 ns will be required. Based on simulation with geant 4 (validated with measurements that used a BGO calorimeter) this will result in an efficiency for the detection of the photons from the positrons annihilation of more than 80\% (arising from the geometrical limitations that do not allow for a full coverage). The measured background for such a signature for the BGO detectors is of the order of 0.25 events/s. By adding the requirement that a positron is detected in the MCP (dark count of 100 counts/s) within 10 ns will result in 0.04 background event/day accidental rate (this includes environmental, internal BGO radioactivity and cosmic rays). The combination of the coincidence between the positron detection in the MCP and the detection of its annihilation photons will result in an efficiency of 64\%.  
The positronium time-of flight will be used to suppress accidentals from the prompt positron and positronium annihilation. 

\subsection{Experimental environment considerations}
Due to its sensitivity to black body radiation the absorber and the mirror have to be kept at 4K. Furthermore, special care has to be taken in order to avoid stray electric and magnetic fields. A mumetal shielding should be foreseen. The target and the electrodes for the Rydberg deceleration can be kept at at higher temperature of 150 K since the collimator will act as a thermal shielding and in addition it will prevent a leak of the electric fields from the deceleration stage. The MCPs are also kept at this higher temperature. 

\section{Expected rate}

The expected signal rate can be estimated using:
\begin{equation}
R_{signal}=R_{e+}\cdot\epsilon_{Ps}\cdot\epsilon_{n33}\cdot\epsilon_{l32}\cdot\epsilon_{collimator}\cdot\epsilon_{tau}\cdot\epsilon_{dec}\cdot\epsilon_{det}=8.7\times10^{-6}\mathrm{events/s}
\end{equation}
where 
\begin{itemize}
\item $R_{e+}=6.25\times 10^6$ e$^+$/s is the positron flux on target. This was estimated assuming that one could use the full intensity of the currently strongest existing positron source, a buffer gas trap efficiency of 20\%, extraction losses to a field free magnetic field region of 70\% and a remoderation efficiency of 5\%. This last one was realized for a continuous positron beam but not yet for a bunched beam as required for this experiment (see Sec. \ref{mod}).   
\item  $\epsilon_{Ps}=0.1$ is the fraction of Ps emitted into vacuum from the proposed porous silica film. This includes the losses due to spin quenching and Ps$_2$ molecule formation and the fact that only the $m=0$ triplet state will be reflected from the proposed mirror.
\item $\epsilon_{n33}=0.25$ is the excitation probability in the $n=33, k=19$ state. 
\item $\epsilon_{l32}=0.1$ the efficiency to drive the Ps in the $n=33, l=32$ state. This has been realized for other atomic species (see e.g. \cite{gallagher} and reference therein) but never for Ps.
\item $\epsilon_{collimator}= 2\times 10^{-9}$ is the fraction of Ps atoms remaining after the collimator ($v_x<100$ m/s and $v_y<0.15$ m/s). This was estimated with a Monte-Carlo simulation (see Sec.\ref{target}).  
\item  $\epsilon_{tau}=0.5$ is the fraction of atoms in the $n=33, l=32$ state, which has a lifetime of about 8 ms, surviving the 5 ms time-of-flight. 
\item  $\epsilon_{dec}$=0.8 the deceleration efficiency for the $v_z<100$ m/s estimated with preliminary simulations validated for different atomic species.
\item $\epsilon_{det}=0.64$ is the efficiency of signal detection. 
\end{itemize}
Assuming that all the efficiencies quoted above could be confirmed experimentally, a rate of 0.7 events/day with a background 0.05 events/day is anticipated. Possible losses due to spurious effects like stray electric or magnetic fields or black body radiation seems to be negligible but further calculations and preliminary experiments should be done to confirm this assumption. 

To note that the expected height of the gravitational state is related to $g$ by equation \ref{eq:z1}. This means that for an uncertainty in the determination of $z_1$ of $dz_1$ one can get an accuracy in the determination of $g$ at the level of $dg/g= 3 dz_1/z_1\sqrt{N}$ where $N$ is the number of detected signals. Assuming an uncertainty of $dz_1 = 0.1$ mm which is mainly determined by the finite source size the value of $dg/g$ can be determined to 3\% in three months. This is comparable to the accuracy that is aimed for anti-hydrogen experiments at CERN \cite{AEGIS, GBAR, ALPHA}. Therefore, observation of Ps gravitational quantum states offers a complementary approach to test the effect of gravity on a pure leptonic system. In comparison to a free fall experiment \cite{Psfreefall} perturbations of the Ps atoms arising from uncontrolled patch elecric fields will not result in a systematic effect for the experiment but will only affect the signal rate.

\section{Conclusions}
The observation of the gravitational quantum states of positronium is very challenging and at this point we cannot conclude that it is feasible. In fact, even though most of the techniques are currently available, some essential milestones have to be demonstrated experimentally before such an experiment could be realized. Those are: the efficient focusing of a bunched positron beam, the Stark deceleration of Rydberg positronium, and the subsequent excitation of Rydberg positronium into states with large angular momentum. In this contribution, we provide our expectations for the efficiencies of these various steps based on published work with Ps and other atomic species. 
Our estimation of the signal rate is encouraging thus justifying and stimulating preliminary experiments, further calculations and simulations. Furthermore, the developed techniques could also find an application in other fields. Rydberg deceleration of Ps atoms would be a major step to open a new era in Ps spectroscopy. If this would be realized the main systematical effects (second order Doppler shift and time-of-flight broadening)  would be reduced by two orders of magnitude (well below its natural linewidth of 1.2 MHz) thus an uncertainty in the measurement of the 1S-2S interval of positronium at the few kHz level might be possible.  Assuming that the QED corrections could be calculated at the same level, this would lead to a new, independent (Ps, being purely leptonic, is free of nuclear size effects) determination of the Rydberg constant at the same level of precision as the current one \cite{CODATA}. This might help to shed some light on the proton size puzzle \cite{muonicH} since this can be solved by shifting by 5 standard deviation the value of the Rydberg constant \cite{nez2011}.

If an observation of the gravitational quantum states of positronium could be realized this would result in a determination of $dg/g$ for positronium at a level of 3\% assuming 70 events could be detected.
If successful, the ongoing efforts to increase the positron buffer gas trap efficiency \cite{dan}, the available slow positron fluxes and seeking to produce a source of colder Ps \cite{brusaPRL,CassidyPRB2012,Consolati}, would certainly help to make the observation of the gravitational quantum states of positronium feasible.

\section*{Acknowledgements}
The authors wish to thank the GRANIT collaboration for providing excellent possibilities for discussions and exchange.\\
P.C. acknowledges the support by the Swiss National Science Foundation (grant PZ00P2 132059) and ETH Zurich (grant ETH-47-12-1) and the very useful discussions with F. Merkt, C. Seiler, D. A. Cooke and D. J. Murtagh.

\end{document}